\documentclass[%
superscriptaddress,
amsmath,amssymb,
aps,
]{revtex4-2}
\usepackage{hyperref}
\usepackage{graphicx}
\usepackage{dcolumn}
\usepackage{bm}
\usepackage{physics}
\usepackage{amsmath}
\usepackage{mathtools}
\usepackage{color}

\begin{document}

	\title{Side-channel-secure quantum key distribution with correlated sources}
	\author{Yang-Guang Shan}
	\author{Jia-Xuan Li}
	\affiliation{Laboratory of Quantum Information, University of Science and Technology of China, Hefei, 230026, Anhui, China}
	\affiliation{Anhui Province Key Laboratory of Quantum Network, University of Science and Technology of China, Hefei, 230026, Anhui, China}
	\affiliation{CAS Center for Excellence in Quantum Information and Quantum Physics, University of Science and Technology of China, Hefei, Anhui 230026, China}
	\author{Zhen-Qiang Yin}
	\email{yinzq@ustc.edu.cn}
	\author{Shuang Wang}
	\author{Wei Chen}
	\author{De-Yong He}
	\author{Guang-Can Guo}
	\author{Zheng-Fu Han}
	\affiliation{Laboratory of Quantum Information, University of Science and Technology of China, Hefei, 230026, Anhui, China}
	\affiliation{Anhui Province Key Laboratory of Quantum Network, University of Science and Technology of China, Hefei, 230026, Anhui, China}
	\affiliation{CAS Center for Excellence in Quantum Information and Quantum Physics, University of Science and Technology of China, Hefei, Anhui 230026, China}
	\affiliation{Hefei National Laboratory, University of Science and Technology of China, Hefei 230088, China}

	\begin{abstract}
		Quantum key distribution (QKD) offers theoretical security guarantees for sharing secure key, but its practical systems face challenges due to the imperfections of devices. Widespread quantum state preparation imperfections, such as correlations between multiple rounds, significantly undermine the real-world security of QKD. In this paper, we propose a protocol that is immune to almost all kinds of state-preparation imperfections over multiple correlated rounds arising from both encoding and unknown non-encoding dimensions. The protocol relies only on three assumptions: the imperfect encoding produces unknown product states rather than entangled ones, a lower bound on the vacuum components is known, and the correlation has a finite range. The proposed protocol is also measurement-device-independent, ensuring high security at both the source and measurement sides. We provide the finite-key security analysis against coherent attacks and conduct numerical simulations to see the performance. The results show that for small correlation ranges, the protocol achieves excellent performance with a maximal transmission loss exceeding 60 dB ($>$300 km in standard fiber). Even for extreme cases, where one encoding affects up to 500 neighboring rounds, the protocol can still generate secret keys over a 10 dB-loss channel.
	\end{abstract}
	\maketitle
\section{Introduction}
Quantum key distribution (QKD) \cite{bennett1984quantum} is a highly practical technology that could help to share secure random numbers between two remote users. In recent years, QKD has made significant advancements both in achieving long distances \cite{wang2022twin,liu2023experimental,liu20231002} and high key rates \cite{li2023high,grunenfelder2023fast}.

Though the security of QKD has been strictly proved in the last few decades \cite{shor2000simple,mayers2001unconditional,renner2008security,tomamichel2012tight,tomamichel2017largely,portmann2022security}, its practical security still remains a nettlesome problem. Realistic devices do not always perform according to the theoretical models, which may open unexpected loopholes for eavesdroppers \cite{RevModPhys.92.025002,brassard2000limitations,lutkenhaus2002quantum,scarani2009security,diamanti2016practical,kang2022patterning}. Device-independent QKD \cite{mayers1998quantum,acin2007device} is a feasible solution to almost all practical loopholes. However, it performs badly both in key rate and transmission distance, and it requires high technical complexity. Luckily, measurement-device-independent (MDI) QKD \cite{braunstein2012side,lo2012measurement} could be immune to all loopholes of the measurement side, maintaining a good performance. Additionally, it can be achieved with current technology at a moderate cost, so it has been extensively researched and has become one of the major protocols of QKD.  

Thanks to MDI QKD, the practical security of QKD improves a lot because most existing attacks are conducted on the measurement side. However, the imperfections of the source side also influence security significantly. The security issues at the source have attracted attention since the early days. For example, the well-known decoy-state method \cite{hwang2003quantum,PhysRevLett.94.230504,wang2005beating} resolved the issue of imperfect single-photon preparation. However, due to the complexity of the source-side devices, ensuring complete security remains challenging.  Realistic devices cannot always prepare the perfect quantum states meeting the requirements of QKD protocols. For example, the intensity modulators do not always output a fixed and predetermined intensity, and the phase modulators may modulate inaccurate and fluctuating phases. What is even more serious is that the modulators may modulate unexpected dimensions of the pulses \cite{PhysRevLett.134.130802}, for example, the time delay or the frequency spectrum of pulses. A few years ago, the side-channel-secure (SCS) QKD \cite{wang2019practical,jiang2023side,jiang2024side} was proposed, which can be immune to almost all imperfections in state preparation. Later, this protocol was extended to the phase-encoding scenario \cite{shan2023practical}. Unluckily, it can only resist uncorrelated imperfections, which means the state of a round cannot be influenced by the encoding choices of other rounds. 

Correlations may come from various reasons \cite{PhysRevA.90.032320,Roberts:18,yoshino2018quantum,curras2023security}, and they can be bi-directional, which means the encoding of a round not only influences the states of its subsequent rounds but also preceding ones. Some existing works trying to resolve the issue of correlation \cite{pereira2020quantum,curras2023security,Zapatero2021securityofquantum,PhysRevApplied.18.044069}. However, existing works require to characterize the correlation in detail, and these works give pessimistic performances. Additionally, there is no security analysis for correlated QKD against general coherent attacks under finite-key case, so these works cannot be directly applied in practical systems. Therefore, existing QKD systems can only rely on the assumption that the correlation is sufficiently small to avoid security issues.

In this work, we present a security proof for a phase-encoding SCS QKD protocol in the presence of state-preparation correlations. The protocol is also MDI. Specifically, the only assumption on the correlations is that they are finite ranged, namely, the encoding choice in the $i$-th round can only affect the quantum states from the $(i-r_1)$-th to the $(i+r_2)$-th rounds, while the overall quantum state remains a product state across different rounds. Notably, our security analysis does not require any characterization of the correlation strength beyond the correlation range specified by $r_1$ and $r_2$. In particular, even an arbitrarily strong correlation, where the encoding bit in one round completely determines the quantum state in another round within the correlation range, does not compromise the security of the protocol. Furthermore, as in other SCS QKD protocols, our analysis requires a lower bound on the vacuum projection probability of the emitted quantum state in each round, without requiring an accurate characterization of the prepared quantum states. For convenience, we refer to this protocol as the finite-correlation-secure (FCS) protocol throughout the remainder of this paper.

We give the finite-key analysis for the FCS protocol against coherent attacks and conduct the numerical simulation to show the performance. Though we do not need to characterize the strength of correlation and only need to know the correlation range, our simulation gives a satisfactory performance. In the case of a small correlation range, the performance is not influenced a lot. Even with a large correlation range, for example, $r_1+r_2=100$, which means each encoding choice could influence the quantum state preparation of up to 100 surrounding rounds, our protocol could also realize a practical transmission distance with attenuation of $30$ dB (corresponding to over 150 km in standard optical fiber). For extreme correlation with a range of 500 pulses, our protocol could still generate secure key under losses exceeding $10$ dB.
	
\section{Results}
We present the results of this study as follows. In the ``Protocol Description'' subsection, we introduce the ideal process of the proposed protocol. Since our protocol does not need the ideal state preparation to ensure security, we introduce the detailed assumptions about state preparation required for the security proof in the ``State Preparation Description'' subsection. Then, in the ``Security Analysis'' subsection, we provide a sketch of the security proof, while a detailed proof can be found in Appendix \ref{sec:appsec}. Finally, in the ``Simulation'' subsection, we present the performance of the proposed protocol.
\subsection{Protocol description}
In the FCS protocol, the two users Alice and Bob act as the source sides, and the untrusted user Charlie acts as the measurement side. The detailed flow is shown below.

\begin{enumerate}
	\item 
	\textbf{State preparation.} In the $i$-th round, Alice (Bob) chooses to select the modulation bit $s_A^i$ ($s_B^i$) from $\{0,1\}$ uniformly at random. Then she (he) tries to prepare the state $\ket{\phi_{s_A}}_{a_i}=\ket{\sqrt{\mu}e^{i\pi s_A^i}}$ ($\ket{\phi_{s_B}}_{b_i}=\ket{\sqrt{\mu}e^{i\pi s_B^i}}$), which is a coherent state with an intensity $\mu$ and a phase $\pi s_A^i$ ($\pi s_B^i$). Then they send the states to the untrusted peer Charlie located in the middle of the channel. We define $s_A=\{s_A^1,s_A^2,\dots\}$ and similar for $s_B$.
	
	Note that in experimental realization, the state $\ket{\phi_{s_{A(B)}}}_{a_i}$ may not only rely on $s_A^i$ ($s_B^i$), but also on $s_A^j$ ($s_B^j$) for some $j\ne i$. And it can have preparation errors and side channels. We will describe the states prepared in detail later.
	\item
	\textbf{State measurement.} If Charlie is honest, the two pulses from Alice and Bob will interfere on a beam splitter. Then Charlie uses two single-photon detectors to detect the two outputs of the interference. We assume the left detector corresponds to the constructive interference and the right detector corresponds to the destructive interference. If only one detector clicks, Charlie will announce a successful click. Charlie also announces which detector clicks. A successful click from the left (right) detector is called a left (right) click and the corresponding round is called a left-click (right-click) round.
	\item 
	\textbf{Sifting.} After $N$ rounds of state preparation and the click announcement of Charlie, $s_A$ and $s_B$ of clicked rounds are kept as sifted key bits. For the right-clicked rounds, Bob should flip his key bit $s_B$.
	\item 
	\textbf{Parameter estimation.} Alice and Bob randomly reveal the key bits of some rounds and count the number of bit errors of these rounds as $n_{\text{est},\text{bit}}$. Each round is independently selected for this parameter estimation with a probability $P_\text{est}$. For the remaining rounds which are not selected for parameter estimation, Alice and Bob count the number of clicks as $n_{\text{sig}}$. If the count $n_{\text{est},\text{bit}}$ is larger than a threshold $n_{\text{est},\text{tol}}$ or the count $n_{\text{sig}}$ is smaller than a threshold $n_{\text{sig},\text{tol}}$, the protocol aborts, otherwise the remaining $n_{\text{sig}}$ bits which are not revealed are kept as raw key bits and the protocol continues.
	\item 
	\textbf{Postprocessing}. Alice and Bob conduct error correction and private amplification to the raw key bits to generate the final secure key.
\end{enumerate}

\subsection{State preparation description}
We have described the ideal states that Alice and Bob want to prepare. However, we do not need an accurate description of the states to keep security. In the following, we will give the assumptions on the state preparation.

Firstly, we still assume uniform and private bit choices, which means $s_A$ and $s_B$ are selected from $\{0,1\}^N$ uniformly at random.

Secondly, we assume only non-entangled correlation exists in state preparation, which means that $\ket{\phi_{s_A}}_{a_i}$ can rely on $s_A^{i-1},s_A^{i+1}$ and so on, but for a given $s_A=\{s_A^1,s_A^2,\dots,s_A^N\}$, the prepared state is shown as a product state $\ket{\phi_{s_A}}_{a_1}\otimes \ket{\phi_{s_A}}_{a_2}\otimes \dots\otimes \ket{\phi_{s_A}}_{a_N}$. Note that this assumption is reasonable because it seems almost impossible to establish entanglement with only laser source, intensity and phase modulators. What is more important, trusted but inaccurate devices correspond to a common scenario.

It is worth noting that, in order for the users to share a common phase reference, this protocol requires a phase-locking technique. However, this does not violate the above assumption. For example, when an optical phase-locked loop is employed, the modulation applied to the light source is determined by the interference measurement outcomes of an auxiliary light. Any inter-round entanglement that may exist in this auxiliary light is not transferred to the prepared quantum states.

\begin{figure}[h]
	\centering
	\includegraphics[width=0.8\textwidth]{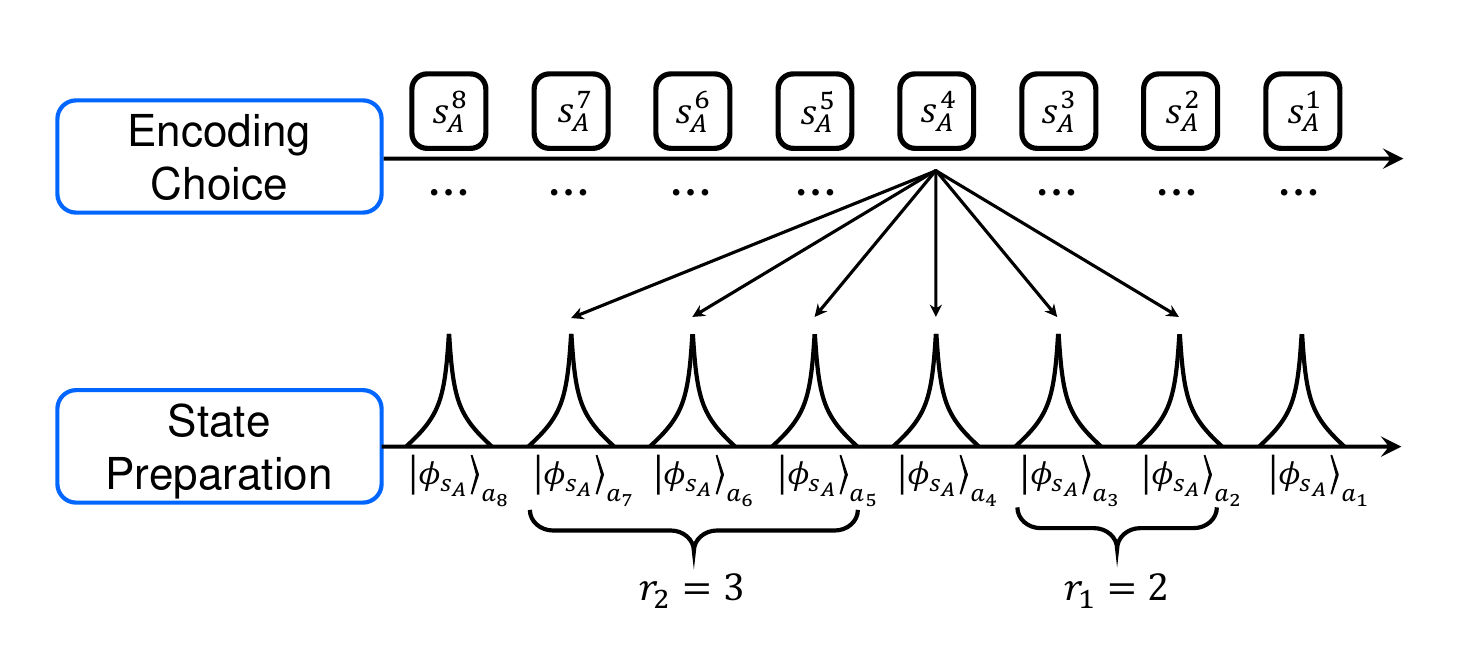}
	\caption{\label{fig:corre}Schematic illustration of correlated state-preparation flaws. The encoding choice made by the sender in each round not only affects the quantum state prepared in that specific round, but may also influence the quantum states prepared in adjacent rounds. For example, in the schematic, the encoding choice in a given round can affect the quantum states in its preceding $r_1=2$ rounds and the subsequent $r_2=3$ rounds. Including the round itself, each encoding bit thus influences the quantum states over a total of $r_1+r_2+1=6$ rounds.}
\end{figure}

Thirdly, we assume the correlation has known finite ranges. The range of backward correlation is defined as $r_1$ and the range of forward correlation is defined as $r_2$, which means $s_A^i$ only influences $\ket{\phi_{s_A}}_{a_{i-r_1}}$ to $\ket{\phi_{s_A}}_{a_{i+r_2}}$. A schematic illustration of the correlation is shown in Fig. \ref{fig:corre}. We may also use $\ket{\phi_{s_A(s_A^j=0)}}_{a_{i}}$ ($\ket{\phi_{s_A(s_A^j=1)}}_{a_{i}}$) when $s_A^j$ is fixed to be $0$ $(1)$ to emphasize the difference between these two states. 

Finally, for the state of each round, we only require a lower bound of the projection probability to a vacuum state. We do not need accurate intensity or phase modulations. This is called side-channel-secure in the SCS QKD \cite{wang2019practical}.

Considering all the assumptions above, we can conclude that in the equivalent protocol, the state prepared by Alice is shown as
\begin{equation}
	\small
	\begin{aligned}
		&\ket{\Phi}_A=\frac{1}{\sqrt{2^{N}}}\sum_{s_A^1,\dots,s_A^N\in\{0,1\}}\bigg(\ket{s_A^1}_{A_1}\otimes\ket{s_A^2}_{A_2}\otimes\dots\ket{s_A^N}_{A_N}\otimes \\
		&\ket{\phi_{s_A}}_{A_pa_1}\otimes\ket{\phi_{s_A}}_{A_pa_2}\otimes \dots\ket{\phi_{s_A}}_{A_pa_i}\otimes\dots\ket{\phi_{s_A}}_{A_pa_N}\bigg),
	\end{aligned}
\end{equation}
where the subscripts $A_1\dots A_N$ correspond to the ancillas held by Alice. The subscripts $a_1\dots a_N$ correspond to the states sent out by Alice. Since these prepared states may not be purified, we use another ancilla with subscript $A_p$ to purify them.

Similar to the SCS QKD \cite{wang2019practical}, we require to know the lower bound of vacuum probability, which means for any $\ket{\phi_{s_A}}_{A_pa_i}$, we have 
\begin{equation}
	\Tr (\ket{\phi_{s_A}}\bra{\phi_{s_A}}_{A_pa_i}\ket{0}\bra{0}_{a_i})\ge\underline{P}_{0A},
\end{equation}
where $\ket{0}$ is the vacuum state. $\underline{P}_{0A}$ is required to be known to Alice and Bob. For simplicity, we also define that 
\begin{equation}
	\ket{\phi_{s_A}}_{A_pa_i}=\sqrt{P_{0A}^i}\ket{\text{puri}_0}_{A_p}\ket{0}_{a_i}+\sqrt{1-P_{0A}^i}\ket{\varphi_1}_{A_pa_i},\label{eq:state}
\end{equation}
for $P_{0A}^i\ge\underline{P}_{0A}$. Note that $P_{0A}^i$ and $\ket{\varphi_1}_{A_pa_i}$ depend on $s_A^j$ for $i-r_2\le j\le i+r_1$, and $\ket{\varphi_1}_{A_pa_i}$ can be an arbitrary state orthogonal to the vacuum $\ket{0}_{a_i}$. Without loss of generality, we assume $\ket{\text{puri}_0}_{Ap}$ to be a fixed normalized state. This assumption can always be satisfied by applying an appropriate isometry to an arbitrary purification. Since any isometry acting on the ancillary system preserves the reduced density operator, the resulting state remains a valid purification of the original quantum state. Therefore, for each prepared state, one can always choose such an isometry so that the ancillary state associated with the vacuum component is the same fixed state $\ket{\text{puri}_0}_{A_p}$. 

The state prepared by Bob is similarly defined as $\ket{\Phi}_B$. We assume that $\ket{\Phi}_A$ and $\ket{\Phi}_B$ are independently prepared as a product state $\ket{\Phi}_A\otimes\ket{\Phi}_B$. Since each round is randomly chosen for parameter estimation with a probability $P_\text{est}$, we introduce a new ancilla to indicate the choice for parameter estimation $\ket{\text{est}}_\text{pe}$ or as a signal round $\ket{\text{sig}}_\text{pe}$. The total state prepared by Alice and Bob is shown as
\begin{equation}
	\begin{aligned}
		\ket{\Phi}=\ket{\Phi}_A\otimes\ket{\Phi}_B\bigotimes_{i=1}^N(\sqrt{P_\text{est}}\ket{\text{est}}_{\text{pe}_i}+\sqrt{1-P_\text{est}}\ket{\text{sig}}_{\text{pe}_i})\\
		\coloneqq\ket{\Phi}_A\otimes\ket{\Phi}_B\bigotimes_{i=1}^N\ket{\psi}_{\text{pe}_i}.
	\end{aligned}
\end{equation}

\subsection{Security proof for FCS QKD}
Because of the complexity, we give the sketch of the security and the result in this section, a detailed analysis can be found in Appendix \ref{sec:appsec}.

Our proof is based on the estimation of phase errors \cite{tomamichel2012tight,tomamichel2017largely}. Alice and Bob prepare the state $\ket{\Phi}$ and acquire their encoding bits by conducting projecting measurements $\{\ket{0}\bra{0},\ket{1}\bra{1}\}$ on their ancillas $A_i$ and $B_i$ for $i\in\{1,2,\dots,N\}$. If we assume that for clicked signal rounds, they measure these ancillas on a complementary basis $\{\ket{+}\bra{+},\ket{-}\bra{-}\}$, where $\ket{\pm}=(\ket{0}\pm\ket{1})/\sqrt{2}$, the error rate could reflect the amount of information leakage to the eavesdropper. The errors of this complementary basis are called phase errors. We call $\{\ket{0},\ket{1}\}$ the $\mathbb{Z}$ basis and $\{\ket{+},\ket{-}\}$ the $\mathbb{X}$ basis. Accordingly, we define a bit error as the event that Alice and Bob obtain the outcomes $\ket{0}\ket{1}$ or $\ket{1}\ket{0}$ ($\ket{0}\ket{0}$ or $\ket{1}\ket{1}$) for left (right) clicks when performing $\mathbb{Z}$-basis measurements on the ancillas. Likewise, we define a phase error as the event that they obtain the outcomes $\ket{+}\ket{+}$ or $\ket{-}\ket{-}$ when performing hypothetical $\mathbb{X}$-basis measurements on the clicked ancillas.

The eavesdropper Eve can have full control over the untrusted party Charlie and perform all operations on Charlie's behalf. Her general coherent attack can be modeled as intercepting all states sent out by the two users, and performing an arbitrary joint measurement, then fabricating Charlie's measurement result for Alice and Bob. We only care the states held by Alice and Bob, so the above description can be expressed as
\begin{equation}
	\rho_{A^NB^N}=\sum_{x}\Tr_{ab}\left(\mathcal{P}\left\{\sqrt{F_x}\ket{\Phi}\ket{LRO_x}^N_C\right\}\right). \label{eq:firststate}
\end{equation}
where $\sum_xF_x=\mathbb{I}$ is a set of POVM on all states $a_i,b_i$ for $i\in\{1,2,\dots,N\}$, representing Eve's general attack. $\mathcal{P}(\ket{\cdot})=\ket{\cdot}\bra{\cdot}$. $\Tr_{ab}$ means the trace operation acting on all $a_i,b_i$ for $i\in\{1,2,\dots,N\}$, $\ket{LRO_x}^N_C$ is a classical register of $N$ trits, representing the measurement result announced by Charlie. It relies on Eve's measurement result $x$, and every trit can be chosen from $\{\ket{L}_C,\ket{R}_C,\ket{O}_C\}$, corresponding to three orthogonal states representing a left click, a right click, or no click.

The central idea of the proof is to analyze the case that the clicked signal rounds of Eq. \ref{eq:firststate} are measured on the $\mathbb{X}$ basis, while the clicked parameter-estimation rounds are measured on the $\mathbb{Z}$ basis. The result of the former corresponds to phase errors, while the result of the later corresponds to $n_{\text{est},\text{bit}}$ in the real implementation. The target of the proof is to relate these values.

We use Kato's concentration inequality \cite{kato2020concentration} to give this relation. This inequality gives the relation between the observations and expectations of random variables, without requirements for independent or identical distributions. Instead, this inequality require some sequential properties. Specifically, in the analysis of our protocol, we assume that Alice and Bob measure the ancillas round by round. Each round may correspond to a clicked parameter-estimation round with a bit error, a clicked signal round with a phase error, or some other events. We define that the measurement outcome of the $u$-th round is $Y_u$, and $\mathcal{F}_u=Y_1Y_2\dots Y_u$. Then, for phase errors, Kato's inequality gives the following relation,
\begin{equation}
	n_{\text{ph}}\le\text{U}_m^{\epsilon^2}(\sum_{u=1}^N\Pr(Y_u=\text{phase error}|\mathcal{F}_{u-1})),
\end{equation}
where $n_{\text{ph}}$ is the number of phase errors in clicked signal rounds, $\Pr(Y_u=\text{phase error}|\mathcal{F}_{u-1})$ is the probability of finding the $u$-th round to be a clicked signal round with a phase error, conditioned on the given results of the first $u-1$ rounds. $\text{U}_m^{\epsilon^2}$ is the function of observation's upper bound, estimated by the conditional expectations with Kato's inequality with a failure probability $\epsilon^2$, which will be given in Appendix \ref{bound}.

With Kato's inequality, the number of bit errors $n_{\text{est},\text{bit}}$ found in parameter-estimation rounds can also be related to its conditional expectation $\Pr(Y_u=\text{bit error, estimation}|\mathcal{F}_{u-1})$, i.e., the conditional probability of finding the $u$-th round to be a parameter-estimation round with a bit error. This relation is shown as:
\begin{equation}
	\sum_{u=1}^N\Pr(Y_u=\text{bit error, estimation}|\mathcal{F}_{u-1})\le \text{U}_e^{\epsilon^2}(n_{\text{est},\text{bit}}),
\end{equation}
where $\text{U}_e^{\epsilon^2}$ is the function of the upper bound of conditional expectations, estimated by the observation with Kato's inequality with a failure probability $\epsilon^2$, which will be given in Appendix \ref{bound}.

Here, the event ``phase error'' corresponds to users' measurement outcomes of $\ket{++}_{A_iB_i}\ket{\text{sig}}_{\text{pe}_i}$ or $\ket{--}_{A_iB_i}\ket{\text{sig}}_{\text{pe}_i}$ for clicked rounds, and the event ``bit error for parameter estimation'' corresponds to users' measurement outcomes of $\ket{01}_{A_iB_i}\ket{\text{est}}_{\text{pe}_i}$ or $\ket{10}_{A_iB_i}\ket{\text{est}}_{\text{pe}_i}$ ($\ket{00}_{A_iB_i}\ket{\text{est}}_{\text{pe}_i}$ or $\ket{11}_{A_iB_i}\ket{\text{est}}_{\text{pe}_i}$) for left (right) clicks. The corresponding terms are shown below.

\begin{equation}
	\begin{aligned}
		&P_{\text{ph}}^u\coloneq\Pr(Y_u=\text{phase error}|\mathcal{F}_{u-1})=\\
		&\frac{\Tr\left\{\otimes_{i= 1}^{u-1}M_i^{AB}(\mathcal{P}(\ket{++}_{A_uB_u})+\mathcal{P}(\ket{--}_{A_uB_u}))\mathcal{P}(\ket{\text{sig}}_{\text{pe}_u})(\ket{L}\bra{L}_{C_u}+\ket{R}\bra{R}_{C_u})\rho_{A^NB^N}\right\}}{\Tr\left\{\otimes_{i= 1}^{u-1}M_i^{AB}\rho_{A^NB^N}\right\}},
	\end{aligned}
\end{equation}

\begin{equation}
	\begin{aligned}
		&P_{\text{est},\text{bit}}^u\coloneq\Pr(Y_u=\text{bit error, estimation}|\mathcal{F}_{u-1})=\\
		&\frac{\Tr\left\{\otimes_{i= 1}^{u-1}M_i^{AB}(\mathcal{P}(\ket{01}_{A_uB_u})+\mathcal{P}(\ket{10}_{A_uB_u}))\mathcal{P}(\ket{\text{est}}_{\text{pe}_u})(\ket{L}\bra{L}_{C_u})\rho_{A^NB^N}\right\}}{\Tr\left\{\otimes_{i= 1}^{u-1}M_i^{AB}\rho_{A^NB^N}\right\}}\\
		+&\frac{\Tr\left\{\otimes_{i= 1}^{u-1}M_i^{AB}(\mathcal{P}(\ket{00}_{A_uB_u})+\mathcal{P}(\ket{11}_{A_uB_u}))\mathcal{P}(\ket{\text{est}}_{\text{pe}_u})(\ket{R}\bra{R}_{C_u})\rho_{A^NB^N}\right\}}{\Tr\left\{\otimes_{i= 1}^{u-1}M_i^{AB}\rho_{A^NB^N}\right\}}.
	\end{aligned}
\end{equation}
Here $\otimes_{i= 1}^{u-1}M_i^{AB}$ represents the measurement for the ancillas of the first $u-1$ rounds. $\mathcal{P}(\ket{\cdot})=\ket{\cdot}\bra{\cdot}$.

To relate $n_{\text{est},\text{bit}}$ and $n_{\text{ph}}$, we find the following relation. A bit error of finding $\ket{01}$ or $\ket{10}$ is equivalent to finding $(\ket{++}-\ket{--})/\sqrt{2}$ or $(\ket{+-}-\ket{-+})/\sqrt{2}$, and finding $\ket{00}$ or $\ket{11}$ is equivalent to finding $(\ket{++}+\ket{--})/\sqrt{2}$ or $(\ket{+-}+\ket{-+})/\sqrt{2}$. By eliminating the terms containing $\ket{++}_{A_uB_u}$ and bounding the terms containing $\ket{+-}$ and $\ket{-+}$ by $0$, we can get the following relation (the detailed derivation is given in Appendix \ref{sec:appsec}):

\begin{equation}
	\begin{aligned}
		\sum_{u=1}^NP_{\text{ph}}^u\le&\frac{2(1-P_\text{est})}{P_\text{est}}\sum_{u=1}^NP_{\text{est},\text{bit}}^u+2(1-P_\text{est})\sum_{u=1}^N\frac{\Tr\left\{\otimes_{i= 1}^{u-1}M_i^{AB}\mathcal{P}(\ket{--}_{A_uB_u})\frac{\mathcal{P}(\ket{\text{sig}}_{\text{pe}_u})}{1-P_\text{est}}\rho_{A^NB^N}\right\}}{\text{const}_u}\\
		&+\frac{2\sqrt{2}(1-P_\text{est})}{\sqrt{P_\text{est}}}\sqrt{\sum_{u=1}^NP_{\text{est},\text{bit}}^u}\sqrt{\sum_{u=1}^N\frac{\Tr\left\{\otimes_{i= 1}^{u-1}M_i^{AB}\mathcal{P}(\ket{--}_{A_uB_u})\frac{\mathcal{P}(\ket{\text{sig}}_{\text{pe}_u})}{1-P_\text{est}}\rho_{A^NB^N}\right\}}{\text{const}_u}},
	\end{aligned}
\end{equation}
where $\text{const}_u=\Tr\left\{\otimes_{i= 1}^{u-1}M_i^{AB}\rho_{A^NB^N}\right\}$. The terms of $P_{\text{est},\text{bit}}^u$ can be estimated by observed error number of parameter-estimation rounds. The rest unknown term is the conditional probability of finding a $\ket{--}_{A_uB_u}$ state, no matter there is a click or not. Since the signal rounds are measured on the $\mathbb{X}$ basis, in this hypothetical complementary measurement, this term can be related to the number of $\ket{--}_{A_iB_i}$ observed in signal rounds with Kato's inequality:
\begin{equation}
	(1-P_\text{est})\sum_{u=1}^N\frac{\Tr\left\{\otimes_{i= 1}^{u-1}M_i^{AB}\mathcal{P}(\ket{--}_{A_uB_u})\frac{\mathcal{P}(\ket{\text{sig}}_{\text{pe}_u})}{1-P_\text{est}}\rho_{A^NB^N}\right\}}{\text{const}_u}\le\text{U}^{\epsilon^2}_e(N_{\text{sig}}^{--}),
\end{equation}
where $N_{\text{sig}}^{--}$ is the number of $\ket{--}_{A_iB_i}$ states observed in both clicked and non-clicked signal rounds . However, $N_{\text{sig}}^{--}$ is still unknown, but it is irrelevant to the eavesdropper's behaviors, because it means the preparing number, not the click number of the $\ket{--}_{A_iB_i}$ states.

Though there are correlations in the state preparation, we can still use the Chernoff bound for independent variables to estimate the upper bound of $N_{\text{sig}}^{--}$ by dividing all rounds into $(r_1+r_2+1)$ groups. Specifically, the $(g+k(r_1+r_2+1))$-th round ($k=0,1,\dots$, and $k\le(N-g)/(r_1+r_2+1)$) is assigned to the $g$-th group. Within each group, the analysis can be carried out under independent conditions, and by combining the results from all groups, the desired upper bound is shown as follows (the detailed derivation is given in Appendix \ref{sec:appsec}):
\begin{equation}
	\begin{gathered}
		\Pr[N_{\text{sig}}^{--}\ge\bar N_{\text{sig}}^{--}]\le(r_1+r_2+1)\epsilon^2,\\
		\bar N_{\text{sig}}^{--}=\sum_{g=1}^{r_1+r_2+1}\text{C}_U^{\epsilon^2}\left[N_g(1-P_\text{est})(1-(\underline{P}_{0A})^{r_1+r_2+1})(1-(\underline{P}_{0B})^{r_1+r_2+1})\right],
	\end{gathered}
\end{equation}
where $N_g$ is the number of rounds in the $g$-th group, $\text{C}_U^{\epsilon^2}$ is the function of upper bound of observations, estimated by the expectations with Chernoff bound, with a failure probability $\epsilon^2$, which will be given in Appendix \ref{bound}.

Summarizing the above derivations, we have completed the phase-error analysis of the FCS protocol. By applying the entropy-uncertainty relation \cite{tomamichel2011uncertainty} and the leftover hash lemma \cite{tomamichel2011leftover}, we can obtain the security parameter $\epsilon_{\text{tot}}$ corresponding to a final key length of $l$. The final results are summarized as follows:
\begin{equation}
	\begin{aligned}
		\bar  n_{\text{ph}}(n_{\text{est},\text{bit}})=&\text{U}_m^{\epsilon^2}\Bigg\{\frac{2(1-P_\text{est})}{P_\text{est}}\text{U}_e^{\epsilon^2}(n_{\text{est},\text{bit}})+2\text{U}^{\epsilon^2}_e\left(N_{\text{sig}}^{--}\right)\\
		&+\frac{2\sqrt{2}\sqrt{1-P_\text{est}}}{\sqrt{P_\text{est}}}\sqrt{\text{U}_e^{\epsilon^2}(n_{\text{est},\text{bit}})}\sqrt{\text{U}^{\epsilon^2}_e\left(N_{\text{sig}}^{--}\right)}\Bigg\},
	\end{aligned}
\end{equation}
\begin{equation}
	l=n_{\text{sig},\text{tol}}(1-H_2(\frac{\bar n_{\text{ph}}(n_{\text{est},\text{tol}})}{n_{\text{sig},\text{tol}}}))-\text{leak}_{ec}-\log_2\frac{2}{\epsilon_{\text{cor}}}-2\log_2\frac{1}{2\tilde \epsilon}.
\end{equation}
\begin{equation}
	\epsilon_{\text{tot}}=\epsilon_{\text{cor}}+2\sqrt{r_1+r_2+4}\epsilon+\tilde\epsilon.
\end{equation}
Here $H_2(x)=-x\log_2(x)-(1-x)\log_2(1-x)$ is the binary Shannon entropy, $\text{leak}_{ec}$ is the number of bits leaked in the error correction step. $\epsilon_{\text{cor}}$ is the correctness parameter and $\tilde{\epsilon}$ is a security parameter introduced in the leftover hash lemma.

\subsection{Numerical simulation}
We conduct numerical simulation to show the performance of the FCS protocol. The device parameters are listed in Table \ref{table:pa}, where $d$ is the dark counting rate per pulse of detectors, $e_{\text{mis}}$ is the misalignment error rate of the interference, $f$ is the efficiency of the error correction step, $\epsilon_{\text{tot}}$ is the total security parameter, and $N$ is the number of rounds conducted by Alice and Bob. 

In the simulation, the number of bits announced in the error correction can be estimated as $\text{leak}_{ec}=fn_{\text{sig}}H_2(e_{\text{bit}})$, where $n_{\text{sig}}$ is the number of clicked signal rounds and $e_{\text{bit}}$ is the bit error rate. We set $\epsilon=\epsilon_{\text{cor}}=\tilde\epsilon=\epsilon_{\text{tot}}/(2+2\sqrt{r_1+r_2+4})$ in the simulation.
\begin{table*}[h]
	\caption{\label{table:pa}The parameters we used in the simulation.}
	\begin{ruledtabular}
	\begin{tabular}{ccccc}
		$d$&$e_{\text{mis}}$&$f$&$\epsilon_{\text{tot}}$&$N$\\[1pt]\colrule
		\noalign{\vskip 1pt}
		$10^{-10}$&$0.01$&$1.1$&$10^{-10}$&$10^{13},10^{14}$\\
	\end{tabular}
\end{ruledtabular}
\end{table*}

\begin{figure}[h]
	\centering
	\includegraphics[width=0.7\textwidth]{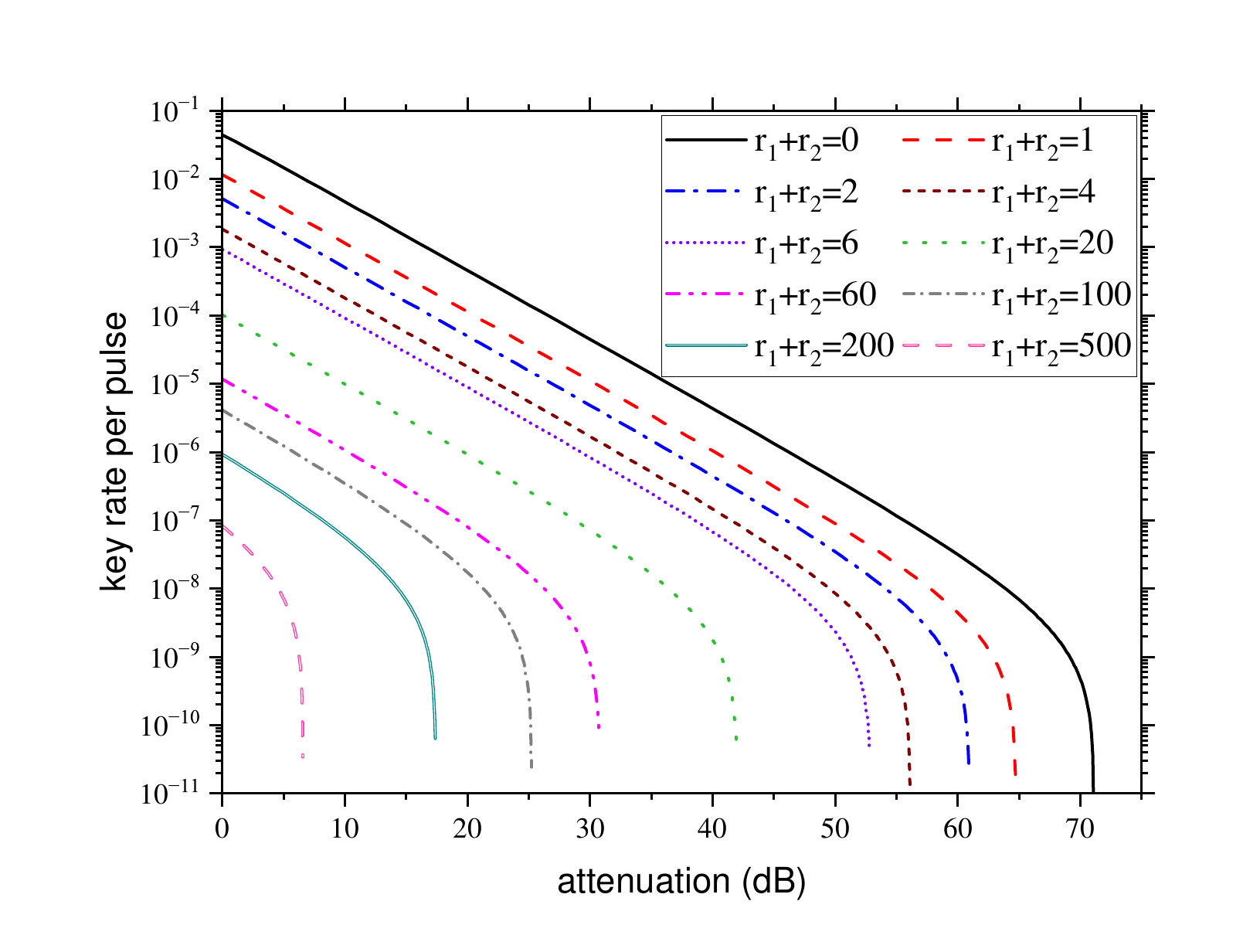}
	\caption{\label{fig:sim13}The simulation result of the FCS protocol under different correlation ranges with $10^{13}$ rounds.}
\end{figure}

\begin{figure}[h]
	\centering
	\includegraphics[width=0.7\textwidth]{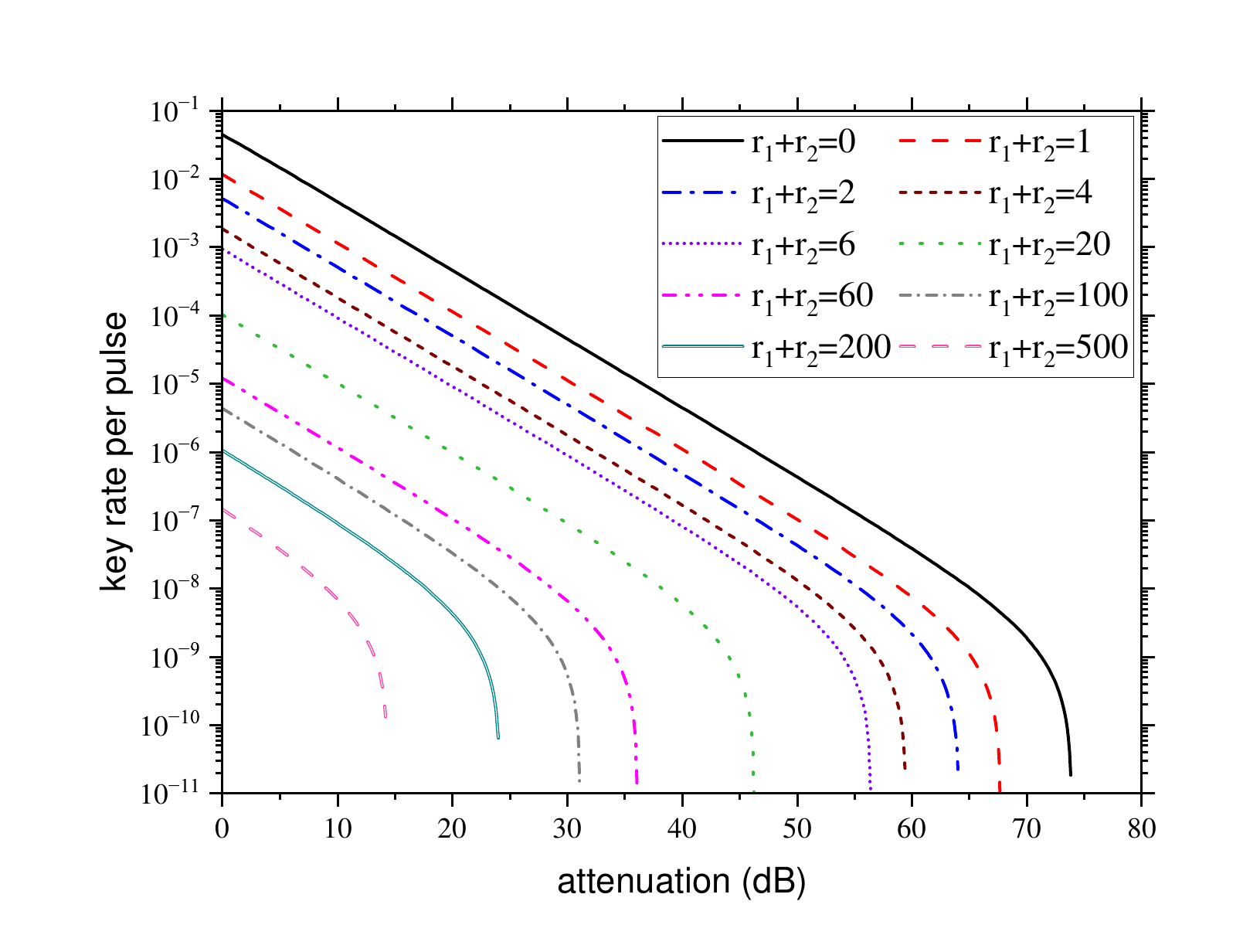}
	\caption{\label{fig:sim}The simulation result of the FCS protocol under different correlation ranges with $10^{14}$ rounds. }
\end{figure}

In the simulation, we assume the correlation and state preparation inaccuracy do not influence the click rates a lot. Thus the key rates are simulated with ideal weak coherent states. Under this condition, the simulation result is only influenced by $r_1+r_2$. We show the results with different $r_1+r_2$ in Fig. \ref{fig:sim13} and Fig. \ref{fig:sim}. Note that the protocol becomes phase-coding side-channel-secure QKD \cite{shan2023practical} when $r_1+r_2=0$.

Fig. \ref{fig:sim13} and Fig. \ref{fig:sim} show the protocol performance for $10^{13}$ and $10^{14}$ executed rounds, respectively. The simulation result shows that a small correlation range does not influence the performance a lot. Compared with the previous works \cite{pereira2020quantum,Zapatero2021securityofquantum,PhysRevApplied.18.044069}, which give pessimistic results under small correlation, our protocol provides a better performance without characterizing the strength of correlation. In addition, when the correlation range is not so large, the performance has been good enough with $10^{13}$ transmitted pulses, demonstrating its practical applicability.

More importantly, our protocol remains practical even when the correlation has a large range. When every single encoding influences the states of $100$ rounds, our protocol could tolerate channel attenuation of more than $30$ dB under finite-key case, corresponding to a fiber channel of more than $150$ km. And when the correlation range reaches an extreme level of $500$ rounds, our protocol can still generate secure key in a channel of $10$ dB loss.

\section{Discussion and conclusion}

The FCS protocol does not require a detailed characterization of the correlation, and only requires the upper bounds of the correlation ranges. This requirement can be easily satisfied because of the following reasons. Firstly, though the issue of correlation is widespread, the correlation range is typically limited. For example, in Ref. \cite{kang2022patterning}, the correlation range is about three, which can be solved by setting a larger range in our FCS protocol. Secondly, if we consider a more general case that the correlation range is unbounded, but correlations beyond a certain range become so weak as to be negligible, our protocol is compatible with the approach of Ref. \cite{pereira2025quantum}, in which correlations outside the threshold can be incorporated as an increase in the security parameter.

For the state prepared in a single round, the FCS protocol only requires to know the lower bound of its projecting probability to the vacuum state. For commonly used weak coherent-state source, the photon-number distribution cannot deviate far from the Poisson distribution. In this case, the users could set a reasonable intensity upper bound or adopt a source monitoring scheme. 

It is worth noting that FCS protocol inherently possesses security against Trojan-horse attacks \cite{vakhitov2001large,gisin2006trojan}. In this kind of attacks, the eavesdropper injects designed light to the source side and analyze the reflected light. The output states can be treated as a superposition of the prepared states of the users and the reflected light of this attack. If the combined quantum states satisfy the assumptions made in the state preparation process, the protocol can still guarantee security.

In terms of the practical implementation of the proposed protocol, the only challenge arises from the requirement for sharing a common phase reference, which has been realized in multiple twin-field QKD experiments. Beyond this, FCS protocol only requires to modulate two phases. Intensity modulation and decoy states are not needed. Therefore, we believe that our protocol provides a practical QKD solution with a high level of practical security.
	\begin{acknowledgments}
		This work has been supported by the National Natural Science Foundation of China ( Grant Nos. 62531023, 62425507, 62271463, 62301524, 62371437 ), 
		the Fundamental Research Funds for the Central Universities (Grant No. WK2030250122), and the Quantum Science and Technology-National Science and Technology Major Project ( Grant No. 2021ZD0300701 ).
		
		\subsection*{Author contributions:}
		Y.-G.S., J.-X.L., and Z.-Q.Y. start this project. Y.-G.S. and Z.-Q.Y. design the protocol. Y.-G.S., J.-X.L., and Z.-Q.Y. provide the framework of the correlation analysis. Y.-G.S. finishes the security analysis for the protocol. Y.-G.S. and Z.-Q.Y. primarily prepare the article. Z.-Q.Y., S.W., W.C., D.-Y.H., G.-C.G., Z.-F.H. supervise this project.
	\end{acknowledgments}
	
	\appendix

\section{Detailed Security Analysis\label{sec:appsec}}
In this section, we give the detailed finite-key security analysis for the FCS protocol against coherent attacks.

We assume a general coherent attack of Eve is shown as a set of POVM $\{F_x\}$ acting on the states $a_i$ and $b_i$ for all $i\in\{1,2,\dots,N\}$ with $\sum_x F_x=\mathbb{I}$, where $\mathbb{I}$ is the identity matrix. For each measurement result corresponding to $F_x$, Eve will pretend to be Charlie and announce a click result $\ket{LRO_x}^N_C$ to Alice and Bob. Here $\ket{LRO_x}^N_C$ is a product state of $N$ sub-state corresponding to a left click $\ket{L}_C$, a right click $\ket{R}_C$ or no click $\ket{O}_C$.

After the announcement of the click result, the state shared by Alice and Bob is shown as 
\begin{equation}
	\sum_{x}\Tr_{ab}\left(\mathcal{P}\left\{\sqrt{F_x}\ket{\Phi}\ket{LRO_x}^N_C\right\}\right),
\end{equation}
where $\Tr_{ab}$ is the trace on the states $a_ib_i$ for all $i\in\{1,2,\dots,N\}$, and $\mathcal{P}\{\ket{\cdot}\}=\ket{\cdot}\bra{\cdot}$.

Our proof is based on the estimation of phase errors \cite{tomamichel2012tight,tomamichel2017largely}. In signal rounds, Alice and Bob will measure their ancillas on the $\mathbb{Z}$ basis ($\ket{0}_{A_i(B_i)}$ and $\ket{1}_{A_i(B_i)}$) to get the key bits. To estimate the phase errors, we need to analyze the case if Alice and Bob measure their ancillas on the $\mathbb{X}$ basis, corresponding to the projection to the states $\ket{+}_{A_i(B_i)}=(\ket{0}_{A_i(B_i)}+\ket{1}_{A_i(B_i)})/\sqrt{2}$ and $\ket{-}_{A_i(B_i)}=(\ket{0}_{A_i(B_i)}-\ket{1}_{A_i(B_i)})/\sqrt{2}$. For both left and right clicks, we define that a phase error corresponds to a measurement result of $\ket{++}_{A_iB_i}$ or $\ket{--}_{A_iB_i}$. This is because for ideal state preparation, $\ket{++}_{A_iB_i}$ and $\ket{--}_{A_iB_i}$ correspond to the cases of even-photon, which are expected to cause few clicks.

In this hypothetical measurement, the signal rounds are measured on the $\mathbb{X}$ basis and the parameter estimation rounds are still measured on the $\mathbb{Z}$ basis. To build the relation between the event numbers of signal and parameter estimation rounds, we assume that Alice and Bob measure the ancillas $A,B$, pe and $C$ round by round.

Before Alice and Bob measure their ancillas of the $u$-th round, they have finished their measurement of the first $u-1$ rounds. At this moment the state (unnormalized) shared by Alice and Bob is shown as 
\begin{equation}
	\sum_{x}\Tr_{ab}\left(\mathcal{P}\left\{(\bigotimes_{i=1}^{u-1}M^{AB}_i)\sqrt{F_x}\ket{\Phi}\ket{LRO_x}^N_C\right\}\right),
\end{equation}
where $M^{AB}_i$ is the measurement matrix of Alice and Bob for their local ancillas of the $i$-th round. For example, if they find that the first round is a left click as a parameter estimation round and they both find a bit 0, we have $M^{AB}_1=\mathcal{P}\left\{\ket{0}_{A_1}\ket{0}_{B_1}\ket{\text{est}}_{\text{pe}_1}\ket{L}_{C_1} \right\}$.

Then the conditional probability of finding a phase error in the $u$-th round is the probability of finding it to be $\ket{++}_{A_uB_u}$ or $\ket{--}_{A_uB_u}$ as a clicked signal round. The corresponding probability is shown as 
\begin{equation}
	\scalebox{0.85}{$\begin{aligned}
			P_{\text{ph}}^u=
			&\frac{
				\sum_{x}\Tr\Big((\ket{++}\bra{++}_{A_uB_u}+\ket{--}\bra{--}_{A_uB_u})(\ket{\text{sig}}_{\text{pe}_u}\bra{\text{sig}}_{\text{pe}_u})(\ket{L}\bra{L}_{C_u}+\ket{R}\bra{R}_{C_u})
				\mathcal{P}\left\{(\bigotimes_{i=1}^{u-1}M^{AB}_i)\sqrt{F_x}\ket{\Phi}\ket{LRO_x}^N_C\right\}\Big)
			}{\sum_{x}\Tr\left(\mathcal{P}\left\{(\bigotimes_{i=1}^{u-1}M^{AB}_i)\sqrt{F_x}\ket{\Phi}\ket{LRO_x}^N_C\right\}\right)}\\
			=&(1-P_\text{est})\frac{\sum_x\Tr\left((\ket{L}\bra{L}_{C_u}+\ket{R}\bra{R}_{C_u})\mathcal{P}\left\{\ket{++}\bra{++}_{A_uB_u}(\bigotimes_{i=1}^{u-1}M^{AB}_i)\sqrt{F_x}\ket{\Phi}_A\ket{\Phi}_B\otimes_{i\ne u}\ket{\psi}_{\text{pe}_i}\ket{LRO_x}^N_C\right\}\right)}{\sum_{x}\Tr\left(\mathcal{P}\left\{(\bigotimes_{i=1}^{u-1}M^{AB}_i)\sqrt{F_x}\ket{\Phi}\ket{LRO_x}^N_C\right\}\right)}\\
			&+(1-P_\text{est})\frac{\sum_x\Tr\left((\ket{L}\bra{L}_{C_u}+\ket{R}\bra{R}_{C_u})\mathcal{P}\left\{\ket{--}\bra{--}_{A_uB_u}(\bigotimes_{i=1}^{u-1}M^{AB}_i)\sqrt{F_x}\ket{\Phi}_A\ket{\Phi}_B\otimes_{i\ne u}\ket{\psi}_{\text{pe}_i}\ket{LRO_x}^N_C\right\}\right)}{\sum_{x}\Tr\left(\mathcal{P}\left\{(\bigotimes_{i=1}^{u-1}M^{AB}_i)\sqrt{F_x}\ket{\Phi}\ket{LRO_x}^N_C\right\}\right)}\\
			\coloneqq&(1-P_\text{est})\frac{\sum_x\Tr((\ket{L}\bra{L}_{C_u}+\ket{R}\bra{R}_{C_u})\left(\mathcal{P}\{\ket{\sigma^{++}_x}_u\}+\mathcal{P}\{\ket{\sigma^{--}_x}_u\}\right))}{\text{const}_u}\\
			=&(1-P_\text{est})\sum_x\frac{\left\Vert\bra{L}_{C_u}\ket{\sigma^{++}_x}_u\right\Vert^2+\left\Vert\bra{L}_{C_u}\ket{\sigma^{--}_x}_u\right\Vert^2+\left\Vert\bra{R}_{C_u}\ket{\sigma^{++}_x}_u\right\Vert^2+\left\Vert\bra{R}_{C_u}\ket{\sigma^{--}_x}_u\right\Vert^2}{\text{const}_u},
		\end{aligned}$}\label{eq:ph}
\end{equation}
where we have define $\ket{\sigma^{++}_x}_u$ and $\ket{\sigma^{--}_x}_u$ to be the huge states in $\mathcal{P}\{\}$ of the second and the third lines, and we also define the denominator for normalization to be $\text{const}_u$.

We need to relate the phase error probability to the observation of parameter estimation rounds. Thus we show the probability of finding the $u$-th round to be a parameter estimation round with a bit error in the following. According to the flow of the protocol, a bit error from a left click corresponds to the measurement result of $\ket{01}_{A_uB_u}$ or $\ket{10}_{A_uB_u}$, and a bit error from a right click corresponds to the measurement result of $\ket{00}_{A_uB_u}$ or $\ket{11}_{A_uB_u}$. Note that finding $\ket{01}$ or $\ket{10}$ is equivalent to finding $(\ket{++}-\ket{--})/\sqrt{2}$ or $(\ket{+-}-\ket{-+})/\sqrt{2}$, and finding $\ket{00}$ or $\ket{11}$ is equivalent to finding $(\ket{++}+\ket{--})/\sqrt{2}$ or $(\ket{+-}+\ket{-+})/\sqrt{2}$. Thus the conditional probability of finding the $u$-th round to be a parameter estimation round with a bit error is shown in the following.
\begin{equation}
	\tiny
	\begin{aligned}
		&P_{\text{est},\text{bit}}^u=\\
		&\frac{
			\sum_{x}\Tr\Big((\ket{\text{est}}_{\text{pe}_u}\bra{\text{est}}_{\text{pe}_u})\Big[\mathcal{P}\{(\ket{++}_{A_uB_u}-\ket{--}_{A_uB_u})/\sqrt{2}\}+\mathcal{P}\{(\ket{+-}_{A_uB_u}-\ket{-+}_{A_uB_u})/\sqrt{2}\}\Big]\ket{L}\bra{L}_{C_u}
			\mathcal{P}\left\{(\bigotimes_{i=1}^{u-1}M^{AB}_i)\sqrt{F_x}\ket{\Phi}\ket{LRO_x}^N_C\right\}\Big)
		}{\sum_{x}\Tr\left(\mathcal{P}\left\{(\bigotimes_{i=1}^{u-1}M^{AB}_i)\sqrt{F_x}\ket{\Phi}\ket{LRO_x}^N_C\right\}\right)}\\
		&+\frac{
			\sum_{x}\Tr\Big((\ket{\text{est}}_{\text{pe}_u}\bra{\text{est}}_{\text{pe}_u})\Big[\mathcal{P}\{(\ket{++}_{A_uB_u}+\ket{--}_{A_uB_u})/\sqrt{2}\}+\mathcal{P}\{(\ket{+-}_{A_uB_u}+\ket{-+}_{A_uB_u})/\sqrt{2}\}\Big]\ket{R}\bra{R}_{C_u}
			\mathcal{P}\left\{(\bigotimes_{i=1}^{u-1}M^{AB}_i)\sqrt{F_x}\ket{\Phi}\ket{LRO_x}^N_C\right\}\Big)
		}{\sum_{x}\Tr\left(\mathcal{P}\left\{(\bigotimes_{i=1}^{u-1}M^{AB}_i)\sqrt{F_x}\ket{\Phi}\ket{LRO_x}^N_C\right\}\right)}\\
		\ge&P_\text{est}\frac{
			\sum_{x}\Tr\Big((\mathcal{P}\{(\ket{++}_{A_uB_u}-\ket{--}_{A_uB_u})/\sqrt{2}\}\ket{L}\bra{L}_{C_u}+\mathcal{P}\{(\ket{++}_{A_uB_u}+\ket{--}_{A_uB_u})/\sqrt{2}\}\ket{R}\bra{R}_{C_u})
			\mathcal{P}\left\{(\bigotimes_{i=1}^{u-1}M^{AB}_i)\sqrt{F_x}\ket{\Phi}\ket{LRO_x}^N_C\right\}\Big)
		}{\sum_{x}\Tr\left(\mathcal{P}\left\{(\bigotimes_{i=1}^{u-1}M^{AB}_i)\sqrt{F_x}\ket{\Phi}\ket{LRO_x}^N_C\right\}\right)}\\
		=&P_\text{est}\sum_x\frac{\left\Vert\bra{L}_{C_u}(\ket{\sigma^{++}_x}_u-\ket{\sigma^{--}_x}_u)/\sqrt{2}\right\Vert^2+\left\Vert\bra{R}_{C_u}(\ket{\sigma^{++}_x}_u+\ket{\sigma^{--}_x}_u)/\sqrt{2}\right\Vert^2}{\text{const}_u}\\
		\ge&P_\text{est}\frac{(\sqrt{\sum_x\left\Vert\bra{L}_{C_u}\ket{\sigma_x^{++}}_u\right\Vert^2}-\sqrt{\sum_x\left\Vert\bra{L}_{C_u}\ket{\sigma_x^{--}}_u\right\Vert^2})^2+(\sqrt{\sum_x\left\Vert\bra{R}_{C_u}\ket{\sigma_x^{++}}_u\right\Vert^2}-\sqrt{\sum_x\left\Vert\bra{R}_{C_u}\ket{\sigma_x^{--}}_u\right\Vert^2})^2}{2\text{const}_u}\\
		\ge&P_\text{est}\frac{(\sqrt{\sum_x\left\Vert\bra{L}_{C_u}\ket{\sigma_x^{++}}_u\right\Vert^2+\sum_x\left\Vert\bra{R}_{C_u}\ket{\sigma_x^{++}}_u\right\Vert^2}-\sqrt{\sum_x\left\Vert\bra{L}_{C_u}\ket{\sigma_x^{--}}_u\right\Vert^2+\sum_x\left\Vert\bra{R}_{C_u}\ket{\sigma_x^{--}}_u\right\Vert^2})^2}{2\text{const}_u},
	\end{aligned}\label{eq:bit}
\end{equation}
where in the first inequality, we discard the terms of $\ket{+-}_{A_uB_u}$ and $\ket{-+}_{A_uB_u}$. In the second inequality, we use the inequality that $\sum_i\Vert\ket{A_i}+\ket{B_i}\Vert^2\ge\abs{\sqrt{\sum_i\Vert\ket{A_i}\Vert^2}-\sqrt{\sum_i\Vert\ket{B_i}\Vert^2}}^2$. In the third inequality, we use the inequality that $(\sqrt{a}-\sqrt{b})^2+(\sqrt{c}-\sqrt{d})^2\ge(\sqrt{a+c}-\sqrt{b+d})^2$. To prove the second inequality, we need to prove that $\sum_i\Vert\ket{A_i}+\ket{B_i}\Vert^2\ge\sum_i\Vert\ket{A_i}\Vert^2+\sum_i\Vert\ket{B_i}\Vert^2-2\sqrt{\sum_i\Vert\ket{A_i}\Vert^2\sum_j\Vert\ket{B_j}\Vert^2}$, which can be proved by using Cauchy-Schwarz inequality for two times:
\begin{equation}
	\begin{aligned}
		&\sum_i\Vert\ket{A_i}\Vert^2+\sum_i\Vert\ket{B_i}\Vert^2-2\sqrt{\sum_i\Vert\ket{A_i}\Vert^2\sum_j\Vert\ket{B_j}\Vert^2}\\
		\le&\sum_i(\Vert\ket{A_i}\Vert^2+\Vert\ket{B_i}\Vert^2-2\Vert\ket{A_i}\Vert\Vert\ket{B_i}\Vert)\\
		\le&\sum_i(\Vert\ket{A_i}\Vert^2+\Vert\ket{B_i}\Vert^2+\braket{A_i}{B_i}+\braket{B_i}{A_i})\\
		=&\sum_i\Vert\ket{A_i}+\ket{B_i}\Vert^2.
	\end{aligned}
\end{equation}

From Eq. (\ref{eq:bit}), we could find that
\begin{equation}
	\sum_x\left\Vert\bra{L}_{C_u}\ket{\sigma_x^{++}}_u\right\Vert^2+\sum_x\left\Vert\bra{R}_{C_u}\ket{\sigma_x^{++}}_u\right\Vert^2\le\left(\sqrt{\frac{2P_{\text{est},\text{bit}}^u\times \text{const}_u}{P_\text{est}}}+\sqrt{\sum_x\left\Vert\bra{L}_{C_u}\ket{\sigma_x^{--}}_u\right\Vert^2+\sum_x\left\Vert\bra{R}_{C_u}\ket{\sigma_x^{--}}_u\right\Vert^2}\right)^2.
\end{equation}
Then substitute it into Eq. (\ref{eq:ph}), we can relate these two probabilities as 
\begin{equation}
	\begin{aligned}
		P_{\text{ph}}^u\le&\frac{2(1-P_\text{est})}{P_\text{est}}P_{\text{est},\text{bit}}^u+2(1-P_\text{est})\frac{\sum_x\left\Vert\bra{L}_{C_u}\ket{\sigma_x^{--}}_u\right\Vert^2+\sum_x\left\Vert\bra{R}_{C_u}\ket{\sigma_x^{--}}_u\right\Vert^2}{\text{const}_u}\\
		&+\frac{2\sqrt{2}(1-P_\text{est})}{\sqrt{P_\text{est}}}\sqrt{P_{\text{est},\text{bit}}^u}\sqrt{\frac{\sum_x\left\Vert\bra{L}_{C_u}\ket{\sigma_x^{--}}_u\right\Vert^2+\sum_x\left\Vert\bra{R}_{C_u}\ket{\sigma_x^{--}}_u\right\Vert^2}{\text{const}_u}}.
	\end{aligned}
\end{equation}

We use Kato's concentration inequality to connect the number of phase errors $n_{\text{ph}}$ and the conditional phase error probabilities of each round $P_{\text{ph}}^u$, which is shown as 
\begin{equation}
	n_{\text{ph}}\le \text{U}_{m}^{\epsilon^2}(\sum_{u=1}^NP_{\text{ph}}^u)\coloneqq\bar{n}_{\text{ph}},\label{eq:nph}
\end{equation}
where $\text{U}_{m}^{\epsilon^2}$ is the upper bound of random variables' observation estimated from their mathematical expectations with a failure probability of $\epsilon^2$, which will be described in detail in Appendix \ref{bound}. $\sum_{u=1}^NP_{\text{ph}}^u$ can be got as follows.
\begin{equation}
	\begin{aligned}
		\sum_{u=1}^NP_{\text{ph}}^u\le&\frac{2(1-P_\text{est})}{P_\text{est}}\sum_{u=1}^NP_{\text{est},\text{bit}}^u+2(1-P_\text{est})\sum_{u=1}^N\frac{\sum_x\left\Vert\bra{L}_{C_u}\ket{\sigma_x^{--}}_u\right\Vert^2+\sum_x\left\Vert\bra{R}_{C_u}\ket{\sigma_x^{--}}_u\right\Vert^2}{\text{const}_u}\\
		&+\frac{2\sqrt{2}(1-P_\text{est})}{\sqrt{P_\text{est}}}\sqrt{\sum_{u=1}^NP_{\text{est},\text{bit}}^u}\sqrt{\sum_{u=1}^N\frac{\sum_x\left\Vert\bra{L}_{C_u}\ket{\sigma_x^{--}}_u\right\Vert^2+\sum_x\left\Vert\bra{R}_{C_u}\ket{\sigma_x^{--}}_u\right\Vert^2}{\text{const}_u}},
	\end{aligned}\label{eq:sumph}
\end{equation}
where we have used the Cauchy-Schwarz inequality that $\sum_i\sqrt{A_iB_i}\le\sqrt{\sum_iA_i}\sqrt{\sum_iB_i}$.

The term $\sum_{u=1}^NP_{\text{est},\text{bit}}^u$ is easy to deal with, which can be related to the number of bit errors in parameter estimation rounds with Kato's inequality:
\begin{equation}
	\sum_{u=1}^NP_{\text{est},\text{bit}}^u\le \text{U}_e^{\epsilon^2}(n_{\text{est},\text{bit}}),\label{eq:nerr}
\end{equation}
where $\text{U}_{e}^{\epsilon^2}$ is the upper bound of random variables' expectation estimated from the observation with a failure probability of $\epsilon^2$, which will be given in detail in Appendix \ref{bound}. $n_{\text{est},\text{bit}}$ is the number of bit errors found in parameter estimation rounds.

Analyzing the second term of Eq. (\ref{eq:sumph}) is a little more difficult. Firstly, we find that $\ket{L}\bra{L}_{C_u}+\ket{R}\bra{R}_{C_u}+\ket{O}\bra{O}_{C_u}=\mathbb{I}$ because either a click (left or right) occurs or no click occurs in each round. Thus we have
\begin{equation} 
	\begin{aligned}
		&\left\Vert\bra{L}_{C_u}\ket{\sigma_x^{--}}_u\right\Vert^2+\left\Vert\bra{R}_{C_u}\ket{\sigma_x^{--}}_u\right\Vert^2\\
		\le&\left\Vert\bra{L}_{C_u}\ket{\sigma_x^{--}}_u\right\Vert^2+\left\Vert\bra{R}_{C_u}\ket{\sigma_x^{--}}_u\right\Vert^2+\left\Vert\bra{O}_{C_u}\ket{\sigma_x^{--}}_u\right\Vert^2\\
		=&\left\Vert\ket{\sigma_x^{--}}_u\right\Vert^2.
	\end{aligned}\label{eq:lrtoall}
\end{equation}
After this scaling, this term becomes the probability that finds a $\ket{--}_{A_uB_u}$ by Alice and Bob, no matter a click occurs or not. Since we are analyzing the case that all signal rounds are measured on the $\mathbb{X}$ basis, we assume that Alice and Bob find $N_{\text{sig}}^{--}$ rounds to be $\ket{--}$ signal events. The probability of finding a signal $\ket{--}_{A_uB_u}$ in the $u$-th round (when they have finished the measurement of the first $u-1$ rounds) is shown as
\begin{equation}
	P_{sig,--}^u=\frac{\sum_x\Tr\left(\ket{\text{sig}}\bra{\text{sig}}_{\text{pe}_u}\otimes\ket{--}\bra{--}_{A_uB_u}\mathcal{P}\left\{(\bigotimes_{i=1}^{u-1}M^{AB}_i)\sqrt{F_x}\ket{\Phi}\ket{LRO_x}^N_C\right\}\right)}{\text{const}_u}=(1-P_\text{est})\frac{\sum_x\left\Vert\ket{\sigma_x^{--}}_u\right\Vert^2}{\text{const}_u}.
\end{equation}
Using Kato's inequality again, we get that 
\begin{equation}
	\sum_{u=1}^NP_{sig,--}^u=(1-P_\text{est})\sum_{u=1}^N\frac{\sum_x\left\Vert\ket{\sigma_x^{--}}_u\right\Vert^2}{\text{const}_u}\le\text{U}^{\epsilon^2}_e(N_{\text{sig}}^{--}).\label{eq:nsig}
\end{equation}
Because we cannot get the number of $N_{\text{sig}}^{--}$ in the realistic experiment, we need to get an upper bound of $N_{\text{sig}}^{--}$. Note that Alice and Bob can measure their ancillas in any order including $A_i,B_i,\text{pe}_i$ and $C_i$ systems, because their measurement matrices on different ancillas commute. We consider the following two kinds of measurements.
\begin{enumerate}
	\item Alice and Bob measure their ancillas round by round, which means before they measure the ancillas $A_u,B_u,\text{pe}_u$ and $C_u$, they have finished the measurement on $A_i,B_i,\text{pe}_i$ and $C_i$ for all $i\in\{1,2,\dots,u-1\}$.
	
	\item Alice and Bob measure the ancillas $\text{pe}_i$, and if they find a $\ket{\text{sig}}_{\text{pe}_i}$, they measure $A_iB_i$ on the $\mathbb{X}$ basis of this round. Note that the order of measurements in different rounds do not influence measurement results. After the above measurement, Alice and Bob measure the rest ancillas.
\end{enumerate}
These two kinds of measurements are equivalent. In the above analysis we use the first kind of measurement, and we will use the second kind to give the upper bound of $N_{\text{sig}}^{--}$. In this measurement, we only care its first step that measures the ancillas $\text{pe}$ and $A,B$, which has been enough to estimate $N_{\text{sig}}^{--}$. The attacks of Eve can be neglected because Alice and Bob have not measured the ancilla $\ket{LRO_x}_C^N$, so we have 
\begin{equation}\begin{aligned}
		&\sum_x\Tr(\sqrt{F_x}^\dagger M_{AB\text{pe}}^\dagger M_{AB\text{pe}}\sqrt{F_x}(\ket{\Phi}\bra{\Phi}\otimes\ket{LRO_x}_C^N\bra{LRO_x}_C^N))\\
		=&\sum_x\Tr(\sqrt{F_x} M_{AB\text{pe}}^\dagger M_{AB\text{pe}}\sqrt{F_x}\ket{\Phi}\bra{\Phi})\\
		=&\Tr(M_{AB\text{pe}}^\dagger M_{AB\text{pe}}(\sum_xF_x)\ket{\Phi}\bra{\Phi})\\
		=&\Tr(M_{AB\text{pe}}^\dagger M_{AB\text{pe}}\ket{\Phi}\bra{\Phi}),
	\end{aligned}
\end{equation}
where $M_{AB\text{pe}}$ is any projection operating on ancillas $A,B$ and $\text{pe}$, and $F_x$ is a POVM operating on the states $a$ and $b$.

To estimate $N_{\text{sig}}^{--}$ in the presence of correlation, we need to group all the rounds. Note that we assumed that the correlation has ranges $r_1$ and $r_2$, which means the encoding of the $u$-th round only influences the states of $(u-r_1)$-th to $(u+r_2)$-th rounds. We divide all the $N$ rounds into $r_1+r_2+1$ groups. The first group contains the $(1+k(r_1+r_2+1))$-th round, the second group contains the $(2+k(r_1+r_2+1))$-th round, and the $g$-th group contains the $(g+k(r_1+r_2+1))$-th round. Here $k=0,1,\dots$. We define $N^{--}_{g}$ to be the number of $\ket{--}$ signal states found in the $g$-th group. Thus we have $N_{\text{sig}}^{--}=\sum_{g=1}^{r_1+r_2+1}N_{g}^{--}$. Though $N^{--}_{g}$ of different $g$ are not independent, we have the following relation:
\begin{equation}
	\begin{aligned}
		\Pr[N_{\text{sig}}^{--}\ge\sum_{g=1}^{r_1+r_2+1}\bar N_{g}^{--}]\le&\Pr[(N^{--}_{1}\ge\bar N^{--}_{1})\lor (N^{--}_{2}\ge\bar N^{--}_{2})\lor\dots\lor (N^{--}_{r_1+r_2+1}\ge\bar N^{--}_{r_1+r_2+1})]\\
		\le&\Pr[N^{--}_{1}\ge\bar N^{--}_{1}]+\Pr[N^{--}_{2}\ge\bar N^{--}_{2}]+\dots+\Pr[N^{--}_{r_1+r_2+1}\ge\bar N^{--}_{r_1+r_2+1}],
	\end{aligned}\label{eq:divide}
\end{equation}
which means that we can estimate the upper bounds $\bar N_{g}^{--}$ of difficult $N_{g}^{--}$ separately, and the total failure probability is bounded by the summation of separate failure probabilities.

Since the order of measurements on different ancillas do not influence the probability distribution of measurement results, we assume other groups have not been measured when we analyze one group. In the following, we will give the analysis of the $g$-th group.

The rounds out of the $g$-th group are not measured in this analysis, so we can take the trace to remove the ancillas $A$ and $B$ which are not in the $g$-th group. However, we do not trace the states $a$ and $b$ in these rounds, because they are correlated with the analyzed rounds. After the trace, the state (unnormalized) becomes
\begin{equation}
	\begin{aligned}
		&(\bigotimes_{i\le(N-g)/(r_1+r_2+1)}\ket{\psi}_{\text{pe}_{g+i(r_1+r_2+1)}})\otimes\\
		&\sum_{\tiny\begin{gathered}s_A^{j}\in\{0,1\}\\[-0.5em]\text{for }j\ne g+i(r_1+r_2+1)\\[-0.5em]i=0,1,\dots
		\end{gathered}}\mathcal{P}\left\{\sum_{\tiny\begin{gathered}s_A^{g+i(r_1+r_2+1)}\in\{0,1\}\\[-0.5em]i=0,1,\dots
		\end{gathered}}\left(\bigotimes_{k\le(N-g)/(r_1+r_2+1)}\ket{s_A^{g+k(r_1+r_2+1)}}_{A_{g+k(r_1+r_2+1)}}\right)\left(\bigotimes_{m=1}^N\ket{\phi_{s_A}}_{A_pa_m}\right)\right\}\otimes\\
		&\sum_{\tiny\begin{gathered}s_B^{j'}\in\{0,1\}\\[-0.5em]\text{for }j'\ne g+i(r_1+r_2+1)\\[-0.5em]i=0,1,\dots
		\end{gathered}}\mathcal{P}\left\{\sum_{\tiny\begin{gathered}s_B^{g+i(r_1+r_2+1)}\in\{0,1\}\\[-0.5em]i=0,1,\dots
		\end{gathered}}\left(\bigotimes_{k\le(N-g)/(r_1+r_2+1)}\ket{s_B^{g+k(r_1+r_2+1)}}_{B_{g+k(r_1+r_2+1)}}\right)\left(\bigotimes_{m=1}^N\ket{\phi_{s_B}}_{B_pb_m}\right)\right\}.
	\end{aligned}\label{eq:stateaftertrace}
\end{equation}

Note that the measurement results of each round belonging to the $g$-th group are not independent. The above state is a mixed state of different $s_A^j\in\{0,1\}$ out of the $g$-th group and the mixing probabilities are uniform. However, if we have got a measurement result of one round, for example, finding a $\ket{+}_{A_{g}}$, the mixing probability may become non-uniform because of the correlation. Thus the measurement results of other rounds can be influenced. 

Though the measurement of different rounds are not independent, we find that they can still be analyzed as if they were independent. We find that the state in Eq. (\ref{eq:stateaftertrace}) is a mixed state of different $s_A^j$ and $s_B^{j'}$ ($j,j'\ne g+i(r_1+r_2+1)$ $(i=0,1,\dots)$). We define $\Pr(s_A^{\sim g},s_B^{\sim g})$ to be the probability of this corresponding term in the mixed state, where $s_{A/B}^{\sim g}$ is $s_{A/B}$ with $s_{A/B}^{g+i(r_1+r_2+1)}$ ($i=0,1,\dots$) removed. Then we have 
\begin{equation}
	\Pr(N^{--}_g\ge \bar N^{--}_g)=\sum_{\tiny\begin{gathered}s_A^{j}\in\{0,1\}\\[-0.5em]\text{for }j\ne g+i(r_1+r_2+1)\\[-0.5em]i=0,1,\dots
	\end{gathered}} \sum_{\tiny\begin{gathered}s_B^{j'}\in\{0,1\}\\[-0.5em]\text{for }j'\ne g+i(r_1+r_2+1)\\[-0.5em]i=0,1,\dots
	\end{gathered}} \Pr(s_A^{\sim g},s_B^{\sim g}) \Pr(N^{--}_g\ge \bar N^{--}_g|s_A^{\sim g},s_B^{\sim g}).\label{eq:condition}
\end{equation}

Now we need to calculate $\Pr(N^{--}_g\ge \bar N^{--}_g|s_A^{\sim g},s_B^{\sim g})$. Conditioned on a fixed $s_A^{\sim g},s_B^{\sim g}$, we will find that the measurement results of different rounds are independent. Under this condition, the state before measurement is shown as 
\begin{equation}
	\begin{aligned}
		&(\bigotimes_{i\le(N-g)/(r_1+r_2+1)}\ket{\psi}_{\text{pe}_{g+i(r_1+r_2+1)}})\otimes\\
		&\mathcal{P}\left\{\sum_{\tiny\begin{gathered}s_A^{g+i(r_1+r_2+1)}\in\{0,1\}\\[-0.5em]i=0,1,\dots
		\end{gathered}}\left(\bigotimes_{k\le(N-g)/(r_1+r_2+1)}\ket{s_A^{g+k(r_1+r_2+1)}}_{A_{g+k(r_1+r_2+1)}}\right)\left(\bigotimes_{m=1}^N\ket{\phi_{s_A}}_{A_pa_m}\right)\right\}_{s_A^{\sim g}}\otimes\\
		&\mathcal{P}\left\{\sum_{\tiny\begin{gathered}s_B^{g+i(r_1+r_2+1)}\in\{0,1\}\\[-0.5em]i=0,1,\dots
		\end{gathered}}\left(\bigotimes_{k\le(N-g)/(r_1+r_2+1)}\ket{s_B^{g+k(r_1+r_2+1)}}_{B_{g+k(r_1+r_2+1)}}\right)\left(\bigotimes_{m=1}^N\ket{\phi_{s_B}}_{B_pb_m}\right)\right\}_{s_B^{\sim g}}.
	\end{aligned}
\end{equation}
Because of the product form of this state, it is obvious that the measurements on $\text{pe}$, on Alice's states and on Bob's states are independent. The probability of finding a $\ket{\text{sig}}$ is $1-P_\text{est}$ for each round. For the measurements on Alice's ancillas, we define that 
\begin{equation}
	\bigotimes_{g+k(r_1+r_2+1)-r_1\le m\le g+k(r_1+r_2+1)+r_2}\ket{\phi_{s_A}}_{A_pa_m}\coloneqq \ket{\phi^{r_1+r_2+1}_{s_A}}_{A_pa_{g,k}}.\label{eq:statemul}
\end{equation}
Then we find that Alice's state is shown as 
\begin{equation}
	\bigotimes_{k\le(N-g)/(r_1+r_2+1)}\mathcal{P}\left\{\sum_{s_A^{g+k(r_1+r_2+1)}\in\{0,1\}}\ket{s_A^{g+k(r_1+r_2+1)}}_{A_{g+k(r_1+r_2+1)}}\ket{\phi_{s_A}^{r_1+r_2+1}}_{A_pa_{g,k}}\right\}_{s_A^{\sim g}}.
\end{equation}
This is because the encoding of the $(g+k(r_1+r_2+1))$-th round $s_A^{g+k(r_1+r_2+1)}$ only influences the states of the neighbor $r_1+r_2+1$ rounds, i.e. $\ket{\phi_{s_A}^{g+k(r_1+r_2+1)}}_{A_pa_{g,k}}$. Because the product form, we find that Alice's measurement on different rounds belonging to the $g$-th group are independent conditioned on a fixed $s_A^{\sim g}$. Additionally, we can obtain the probability of finding a $\ket{-}_{g+k(r_1+r_2+1)}$ in the following.

The state related to the measurement of the $(g+k(r_1+r_2+1))$-th round is shown as 
\begin{equation}
	\left(\ket{0}_{A_{g+k(r_1+r_2+1)}}\ket{\phi_{s_A(s_A^{g+k(r_1+r_2+1)}=0)}^{r_1+r_2+1}}_{A_pa_{g,k}}+\ket{1}_{A_{g+k(r_1+r_2+1)}}\ket{\phi_{s_A(s_A^{g+k(r_1+r_2+1)}=1)}^{r_1+r_2+1}}_{A_pa_{g,k}}\right)/\sqrt{2}.
\end{equation}
Then the probability of finding a $\ket{-}_{A_{g+k(r_1+r_2+1)}}$ is shown as 
\begin{equation}
	P^{-}_A=\frac{1}{4}\left\Vert\ket{\phi_{s_A(s_A^{g+k(r_1+r_2+1)}=0)}^{r_1+r_2+1}}_{A_pa_{g,k}}-\ket{\phi_{s_A(s_A^{g+k(r_1+r_2+1)}=1)}^{r_1+r_2+1}}_{A_pa_{g,k}}\right\Vert^2.
\end{equation}
Recall the assumption of the states (in the \textbf{State preparation description} section of the main text) and the definition in Eq. (\ref{eq:statemul}), the states above can be expressed as 
\begin{equation}
	\begin{aligned}
		\ket{\phi_{s_A(s_A^{g+k(r_1+r_2+1)}=0)}^{r_1+r_2+1}}_{A_pa_{g,k}}=\sqrt{\prod_{i=g+u(r_1+r_2+1)-r_1}^{g+u(r_1+r_2+1)+r_2} P_{0A}^i}(\ket{\text{puri}_0}_{Ap}\ket{0}_{a})^{\otimes (r_1+r_2+1)}+\sqrt{1-\prod_{i=g+u(r_1+r_2+1)-r_1}^{g+u(r_1+r_2+1)+r_2} P_{0A}^i}\ket{\varphi_0^u}_{A_pa_{g,k}},\\
		\ket{\phi_{s_A(s_A^{g+k(r_1+r_2+1)}=1)}^{r_1+r_2+1}}_{A_pa_{g,k}}=\sqrt{\prod_{i=g+u(r_1+r_2+1)-r_1}^{g+u(r_1+r_2+1)+r_2} P_{0A}^{i\prime}}(\ket{\text{puri}_0}_{Ap}\ket{0}_{a})^{\otimes (r_1+r_2+1)}+\sqrt{1-\prod_{i=g+u(r_1+r_2+1)-r_1}^{g+u(r_1+r_2+1)+r_2} P_{0A}^{i\prime}}\ket{\varphi_1^u}_{A_pa_{g,k}},
	\end{aligned}
\end{equation}
where $\ket{\varphi_{0/1}^u}_{A_pa_{g,k}}$ is orthogonal to the state $\ket{0}_a^{\otimes(r_1+r_2+1)}$. We assume that $\braket{\varphi_{0}^u}{\varphi_{1}^u}=X_A$. Then we find that 
\begin{equation}
	\scalebox{0.8}{$
		\begin{aligned}
			P^{-}_A=&\frac{1}{4}\left(2-2\sqrt{\prod_{i=g+u(r_1+r_2+1)-r_1}^{g+u(r_1+r_2+1)+r_2} P_{0A}^i}\sqrt{\prod_{i=g+u(r_1+r_2+1)-r_1}^{g+u(r_1+r_2+1)+r_2} P_{0A}^{i\prime}}-\sqrt{1-\prod_{i=g+u(r_1+r_2+1)-r_1}^{g+u(r_1+r_2+1)+r_2} P_{0A}^i}\sqrt{1-\prod_{i=g+u(r_1+r_2+1)-r_1}^{g+u(r_1+r_2+1)+r_2} P_{0A}^{i\prime}}(X_A+X_A^*)\right)\\
			\le&\frac{1}{4}\left(2-2(\underline{P}_{0A})^{r_1+r_2+1}+2(1-(\underline{P}_{0A})^{r_1+r_2+1})\right)\\
			=&(1-(\underline{P}_{0A})^{r_1+r_2+1}).
		\end{aligned}$}
\end{equation}
It is easy to find the upper bound is obtained when $X_A=-1$. Then it becomes monotonically decreasing for all $P_{0A}^i$ and $P_{0A}^{\prime i}$. 

The analysis of Bob's state is similar. Thus conditioned on fixed $s_A^{\sim g}, s_B^{\sim g}$, the probability of finding a $\ket{\text{sig}}_{\text{pe}_{g+k(r_1+r_2+1)}}\otimes\ket{--}_{A_{g+k(r_1+r_2+1)}B_{g+k(r_1+r_2+1)}}$ is shown as $(1-P_\text{est})(1-(\underline{P}_{0A})^{r_1+r_2+1})(1-(\underline{P}_{0B})^{r_1+r_2+1})$. We use the Chernoff bound for independent random variables to get the following relations:
\begin{equation}
	\begin{gathered}
		\Pr(N^{--}_g\ge \bar N^{--}_g|s_A^{\sim g},s_B^{\sim g})\le\epsilon^2,\\
		\bar N^{--}_g=\text{C}_U^{\epsilon^2}\left[N_g(1-P_\text{est})(1-(\underline{P}_{0A})^{r_1+r_2+1})(1-(\underline{P}_{0B})^{r_1+r_2+1})\right],
	\end{gathered}\label{eq:independent}
\end{equation}
where $N_g$ is the number of rounds in the $g$th group, and $\text{C}_U^{\epsilon^2}$ is the upper bound of measurement result of random variables estimated from their expectations with Chernoff bound. The details of Chernoff bound is shown in Appendix \ref{bound}. Note that the above relations hold for any $s_A^{\sim g}$ and $s_B^{\sim g}$, so Eq. (\ref{eq:condition}) becomes
\begin{equation}
	\Pr(N^{--}_g\ge \bar N^{--}_g)\le \epsilon^2,
\end{equation}
for the given $\bar N^{--}_g$ in Eq. (\ref{eq:independent}).

Substitute these results into Eq. (\ref{eq:divide}), we get that 
\begin{equation}
	\begin{gathered}
		\Pr[N_{\text{sig}}^{--}\ge\bar N_{\text{sig}}^{--}]\le(r_1+r_2+1)\epsilon^2,\\
		\bar N_{\text{sig}}^{--}=\sum_{g=1}^{r_1+r_2+1}\bar N_g^{--}.
	\end{gathered}\label{eq:nsigest}
\end{equation}

Finally, combining Eqs. (\ref{eq:nph},\ref{eq:sumph},\ref{eq:nerr},\ref{eq:lrtoall},\ref{eq:nsig},\ref{eq:independent},\ref{eq:nsigest}), we get the final upper bound of phase errors:
\begin{equation}
	\begin{aligned}
		\bar  n_{\text{ph}}(n_{\text{est},\text{bit}})=&\text{U}_m^{\epsilon^2}\Bigg\{\frac{2(1-P_\text{est})}{P_\text{est}}\text{U}_e^{\epsilon^2}(n_{\text{est},\text{bit}})+2\text{U}^{\epsilon^2}_e\left(\sum_{g=1}^{r_1+r_2+1}\text{C}_U^{\epsilon^2}\left[N_g(1-P_\text{est})(1-(\underline{P}_{0A})^{r_1+r_2+1})(1-(\underline{P}_{0B})^{r_1+r_2+1})\right]\right)\\
		&+\frac{2\sqrt{2}\sqrt{1-P_\text{est}}}{\sqrt{P_\text{est}}}\sqrt{\text{U}_e^{\epsilon^2}(n_{\text{est},\text{bit}})}\sqrt{\text{U}^{\epsilon^2}_e\left(\sum_{g=1}^{r_1+r_2+1}\text{C}_U^{\epsilon^2}\left[N_g(1-P_\text{est})(1-(\underline{P}_{0A})^{r_1+r_2+1})(1-(\underline{P}_{0B})^{r_1+r_2+1})\right]\right)}\Bigg\},
	\end{aligned}
\end{equation}
the corresponding failure probability is $(r_1+r_2+1)\epsilon^2+3\epsilon^2=(r_1+r_2+4)\epsilon^2$.

For the sub-normalized quantum state after the passing of the parameter estimation and error verification, its upper bound of $n_{\text{ph}}$ is $\bar n_{\text{ph}}(n_{\text{est},\text{tol}})$ with a failure probability less than $(r_1+r_2+4)\epsilon^2$ \cite{tomamichel2017largely}. In the framework of composable security \cite{ben2004general,unruh2004simulatable,muller2009composability}, using the leftover hashing lemma \cite{renner2008security,tomamichel2011leftover,tomamichel2015quantum,tomamichel2017largely}, the secrecy parameter is shown as $\epsilon_\text{sec}=2\epsilon'+\frac{1}{2}\sqrt{2^{l-H_\text{min}^{\epsilon'}(Z|E')}}$, where $Z$ corresponds to Alice's key bits (measurement results of her ancillas on $\mathbb{Z}$ basis) and $E'$ is Eve's system and $l$ is the length of the final key. $E'$ can be divided into two parts, the classical information from the error correction step $E_{ec}$ announced by the two users and other states $E$ held by Eve. From the property of smooth entropy (Proposition 5.10 of \cite{tomamichel2012framework}), we have 
\begin{equation}
	H_\text{min}^{\epsilon'}(Z|E')\ge H_\text{min}^{\epsilon'}(Z|E)-\log_2(\dim(E_{ec}))=H_\text{min}^{\epsilon'}(Z|E)-\text{leak}_{ec}-\lceil\log_2\frac{1}{\epsilon_{\text{cor}}}\rceil,
\end{equation}
where $\text{leak}_{ec}$ is the number of bits leaked in the error correction and $\lceil\log_2\frac{1}{\epsilon_{\text{cor}}}\rceil$ is the number of bits leaked in the error verification.

From the uncertainty relation of smooth entropies \cite{tomamichel2011uncertainty}, we have 
\begin{equation}
	H_\text{min}^{\epsilon'}(Z|E)\ge n_{\text{sig},\text{tol}}-H_\text{max}^{\epsilon'}(X|E)\ge n_{\text{sig},\text{tol}}(1-H_2(\frac{\bar n_{\text{ph}}(n_{\text{est},\text{tol}})}{n_{\text{sig},\text{tol}}})),
\end{equation}
with $\epsilon'=\sqrt{r_1+r_2+4}\epsilon$ \cite{tomamichel2017largely}. Here we have use the requirement of passing the parameter estimation that $n_{\text{sig}}\ge n_{\text{sig},\text{tol}}$.

We can set $l=H_\text{min}^{\epsilon'}(Z|E')-2\log_2\frac{1}{2\tilde \epsilon}$. Thus we have $\epsilon_\text{sec}=2\epsilon'+\tilde \epsilon=2\sqrt{r_1+r_2+4}\epsilon+\tilde\epsilon$.

Considering the error verification step announcing $\lceil \log_2(1/\epsilon_{\text{cor}})\rceil$ bits fails with a probability $\epsilon_{\text{cor}}$, the total security parameter is shown as $\epsilon_{\text{tot}}=\epsilon_{\text{cor}}+\epsilon_\text{sec}=\epsilon_{\text{cor}}+2\sqrt{r_1+r_2+4}\epsilon+\tilde\epsilon$. The final key length is shown as 
\begin{equation}
	l=n_{\text{sig},\text{tol}}(1-H_2(\frac{\bar n_{\text{ph}}(n_{\text{est},\text{tol}})}{n_{\text{sig},\text{tol}}}))-\text{leak}_{ec}-\log_2\frac{2}{\epsilon_{\text{cor}}}-2\log_2\frac{1}{2\tilde \epsilon}.
\end{equation}

\section{Kato's inequality and Chernoff bound\label{bound}}
In this section, we will simply introduce Kato's inequality and Chernoff bound used in our analysis.
\subsection{Kato's inequality}
Kato's inequality \cite{kato2020concentration} is an improved version of Azuma's inequality \cite{azuma1967weighted}, which has been widely used in the security analysis of quantum key distribution.
\\

\textbf{Kato's inequality.} Let $\{X_m\}$ be a list of random variables, and $\mathcal{F}_m$ be the measurement result of the random variables $X_1,X_2,\dots,X_m$. Suppose that $0\le X_m\le1$ for all m. In this case, for any $n\in \mathbb{N}$, $a\in\mathbb{R}$ and $b\in\mathbb{R}_{\ge0}$,
\begin{equation}
	P\left(\sum_{m=1}^n(E(X_m|\mathcal{F}_{m-1})-X_m)\ge(b+a(2\frac{\sum_{m=1}^n X_m}{n}-1))\sqrt{n}\right)\le\exp(-\frac{2(b^2-a^2)}{(1+\frac{4a}{3\sqrt{n}})^2})
\end{equation}
holds.
\\

We denote that $\Lambda=\sum_m^nX_m$, then we can get that
\begin{equation}
	\text{Pr}\left(\sum_{m=1}^nE(X_m|\mathcal{F}_{m-1})-\Lambda\ge\left(b+a(\frac{2\Lambda}{n}-1)\right)\sqrt{n}\right)\le\exp\left(-\frac{2(b^2-a^2)}{(1+\frac{4a}{3\sqrt{n}})^2}\right)\label{concen:1},
\end{equation}
and
\begin{equation}
	\text{Pr}\left(\Lambda-\sum_{m=1}^nE(X_m|\mathcal{F}_{m-1})\ge\left(b+a(\frac{2\Lambda}{n}-1)\right)\sqrt{n}\right)\le\exp\left(-\frac{2(b^2-a^2)}{(1-\frac{4a}{3\sqrt{n}})^2}\right)\label{concen:2},
\end{equation}
by replacing $X_m\to 1-X_m$ and $a\to -a$. \cite{curras2021tight}

For Eq. (\ref{concen:1}), to get a tight bound, we let $\exp(-\frac{2(b^2-a^2)}{(1+\frac{4a}{3\sqrt{n}})^2})=\epsilon$ and solve $\min[\left(b+a(\frac{2\Lambda}{n}-1)\right)]$. The optimal value of $a$ and $b$ are
\begin{equation}
	a_1=\frac{3\left(72\sqrt{n}\Lambda(n-\Lambda)\ln\epsilon-16n^{3/2}\ln^2\epsilon+9\sqrt{2}(n-2\Lambda)\sqrt{-n^2\ln\epsilon(9\Lambda(n-\Lambda)-2n\ln\epsilon)}\right)}{4(9n-8\ln\epsilon)(9\Lambda(n-\Lambda)-2n\ln\epsilon)},
\end{equation}
\begin{equation}
	b_1=\frac{\sqrt{18a^2n-(16a^2+24a\sqrt{n}+9n)\ln\epsilon}}{3\sqrt{2n}}.
\end{equation}

Then for known measurement result of random variables (known $\Lambda$), the upper bound of mathematical expectations can be get by 
\begin{equation}
	\sum_{m=1}^nE(X_m|\mathcal{F}_{m-1})\le\Lambda+\left(b_1+a_1(\frac{2\Lambda}{n}-1)\right)\sqrt{n}\equiv \text{U}_e^\epsilon(\Lambda).
\end{equation}
And with known expectations, the lower bound of $\Lambda$ is 
\begin{equation}
	\Lambda\ge\frac{\sum_{m=1}^nE(X_m|\mathcal{F}_{m-1})-(b_1-a_1)\sqrt{n}}{1+\frac{2a_1}{\sqrt{n}}}\equiv \text{L}_m^\epsilon(\sum_{m=1}^nE(X_m|\mathcal{F}_{m-1})).
\end{equation}

For Eq. (\ref{concen:2}), with a similar process, we can get 
\begin{equation}
	a_2=\frac{-3\left(72\sqrt{n}\Lambda(n-\Lambda)\ln\epsilon-16n^{3/2}\ln^2\epsilon-9\sqrt{2}(n-2\Lambda)\sqrt{-n^2\ln\epsilon(9\Lambda(n-\Lambda)-2n\ln\epsilon)}\right)}{4(9n-8\ln\epsilon)(9\Lambda(n-\Lambda)-2n\ln\epsilon)},
\end{equation}
\begin{equation}
	b_2=\frac{\sqrt{18a^2n-(16a^2-24a\sqrt{n}+9n)\ln\epsilon}}{3\sqrt{2n}},
\end{equation}

\begin{equation}
	\sum_{m=1}^nE(X_m|\mathcal{F}_{m-1})\ge\Lambda-\left(b_2+a_2(\frac{2\Lambda}{n}-1)\right)\sqrt{n}\equiv \text{L}_e^\epsilon(\Lambda),
\end{equation}

\begin{equation}
	\Lambda\le\frac{\sum_{m=1}^nE(X_m|\mathcal{F}_{m-1})+(b_2-a_2)\sqrt{n}}{1-\frac{2a_2}{\sqrt{n}}}\equiv \text{U}_m^\epsilon(\sum_{m=1}^nE(X_m|\mathcal{F}_{m-1})).
\end{equation}

\subsection{Chernoff bound}
In our security analysis, we use the Chernoff bound \cite{mitzenmacher2017probability} to get the upper bound of measurement result for independent random variables.
\\

\textbf{Multiplicative Chernoff bound.} Suppose $X_1,X_2,\dots,X_n$ are independent random variables taking values in $\{0,1\}$. Let $X=X_1+X_2+\dots+X_n$ and $\mu=E[X]$ is its expectation. Then for $\delta>0$,
\begin{equation}
	P(X>(1+\delta)\mu)\le(\frac{\text{e}^{\delta}}{(1+\delta)^{1+\delta}})^\mu,
\end{equation}
and
\begin{equation}
	P(X<(1-\delta)\mu)\le(\frac{\text{e}^{-\delta}}{(1-\delta)^{1-\delta}})^\mu
\end{equation}
\\

With the equality that $\frac{2\delta}{2+\delta}\le\ln(1+\delta)$, we can get the inequality we used.
\begin{equation}
	P(X\ge(1+\delta)\mu)\le\text{e}^{-\delta^2\mu/(2+\delta)},\ \ \delta\ge0.
\end{equation}
We let $\text{e}^{-\delta^2\mu/(2+\delta)}=\epsilon$, then
\begin{equation}
	\text{C}_U^\epsilon(\mu)\equiv (1+\delta)\mu
\end{equation}
\begin{equation}
	\delta=\frac{\ln\frac{1}{\epsilon}+\sqrt{(\ln\frac{1}{\epsilon})^2+8\mu\ln\frac{1}{\epsilon}}}{2\mu}
\end{equation}
	
	\bibliography{refe.bib}
\end{document}